\documentclass[%twocolumn,
aps,prd,  
               preprintnumbers,numbers,sort&compress,
               nofootinbib, showpacs, colorlinks,
               linkcolor=blue, citecolor=blue]{revtex4-2} % reprint,nofootinbib

\usepackage{graphicx}% Include figure files
\usepackage{dcolumn}% Align table columns on decimal point
\usepackage{bm}% bold math
\usepackage{hyperref}
\usepackage{xcolor}

\usepackage{amsmath}
\usepackage[caption=false]{subfig}
\usepackage{gensymb}% to type the \degree symbol
\usepackage{amsmath}
 \usepackage{bm}% bold math
\usepackage{graphicx}% Include figure files
\newcommand{\be}{\begin{eqnarray}}
\newcommand{\ee}{\end{eqnarray}}
\newcommand{\beq}{\begin{equation}}
\newcommand{\eeq}{\end{equation}}
\newcommand{\exclude}[1]{}

\def\ra{\rangle}
\def\la{\langle}

 \newcommand{\jcap}{JCAP}
 \graphicspath{{./Figures/}} 

\begin{document}
    
       \title{Why   supersonic UAP events  are  not accompanied by   acoustic booms?  }
        \author{Ariel  Zhitnitsky  }
               \email{arz@phas.ubc.ca}
       \affiliation{  University of British Columbia,Vancouver,  Canada}

     \begin{abstract}
     
     {The Unidentified Anomalous Phenomena (UAP) events have been recorded for decades. However, the nature of these events remains unknown. 
     We advocate an idea that  the dark matter  (DM) in form of the     axion quark nuggets (AQN)    made of standard model  quarks (or {\it antiquarks}) and gluons could 
{\it simultaneously} explain  many  observed puzzling features which always accompany the UAP events, such as   high velocity, enormous acceleration, absence   of any remaining materials in affected area, etc. This long list of mysterious  features  also includes the absence of any sound which normally accompanies any supersonic motion. The AQN framework answers this  last puzzle as well, which represents the title of the present work.    The AQN dark matter model was invented long ago without any intension  to apply the model to UAP events.  Rather,   it was invented with  a single motivation to explain the observed  similarity     $\Omega_{\rm DM}\sim \Omega_{\rm visible}$ between  visible and DM components, which  represents a very  generic feature of this framework,  not sensitive  to any parameters of the construction. In this work we use the same framework with the same set of parameters to study the observed UAP events. Essentially we argue that a specific  class of UAP   represents the  close cousin of 
other mysteries known for decades (even centuries), such as  sky-quakes, Ball Lightnings, and Mysterious Transient  events. In other words, the UAPs  represent different manifestations of exactly the same objects, the AQNs propagating in atmosphere. The fuel powering all these mysterious energetic events is one and the same: it is the antiquark core of the AQNs. 
We also present   some  suggestions on   type of instruments required to  study this specific (and well defined) type  of the   UAP events.}
 
           \end{abstract}

	% \begin{keywords}
%	 dark matter,    axion quark nugget,  mysterious transients, ball lighting, sky-quakes, UAP  
%  \end{keywords}

\maketitle

\section{Introduction and Motivation}\label{sec:introduction}
This work represents a continuation of our previous papers \cite{Budker:2020mqk,Zhitnitsky:2025bvy,Zhitnitsky:2026pai} devoted to explanation of  some mysterious atmospheric events. 
%which can be classified (in modern terminology)  as  Unidentified Anomalous Phenomena (UAP) events.   
More specifically,  these papers  study the so-called sky-quakes  \cite{Budker:2020mqk},  Ball Lightning (BL) events \cite{Zhitnitsky:2025bvy}, known for centuries,  and Mysterious Transient (MT)  events  \cite{Zhitnitsky:2026pai} as   identified by the  Palomar Observatory
Sky Survey (POSS-1).
% Numerous (but unsuccessful) attempts had been previously made to understand these puzzling events in terms of conventional  physics.  
All the analysis  for these dramatically different manifestations of the puzzling events in atmosphere were based on one and the same specific dark matter (DM) model,   the so-called Axion Quark Nugget (AQN) model 
when  the   DM object   is represented by composite macroscopically large AQNs   made of the  standard model  quarks (or antiquarks) and gluons. It should be  contrasted with conventional 40 years old DM paradigm  when   DM particles   are  represented by some fundamental new (unknown and not observed) fields  of BSM (Beyond Standard Model) physics. The AQN   behaves as a chameleon: it serves as a proper DM object in dilute galactic environment, but becomes very strongly interacting (with surrounding material) object 
 when it hits the stars or planets. As a result these objects  may produce   profound and very energetic events. 
 
 The main goal of the present work is to argue that the same AQNs may explain a number of profound  Unidentified Anomalous Phenomena (UAP) events within the same framework with the same set of parameters being used in  \cite{Budker:2020mqk,Zhitnitsky:2025bvy,Zhitnitsky:2026pai}. Essentially we argue that a specific  class of UAP events represent the  close cousins of the sky-quakes, BL, and MT events  discussed in \cite{Budker:2020mqk,Zhitnitsky:2025bvy,Zhitnitsky:2026pai}, and these UAPs  represent different manifestations of exactly the same objects, the AQNs propagating in atmosphere.

 \exclude{
 %as reported by professor Avi Loeb \cite{DOW-UAP-D108} and possible interpretation as 
 \cite{haqqmisra2026objectsreported480mph}
 The basic features of the UAP events   reported in \cite{DOW-UAP-D108} resembles very much the properties of UAPs as previously discussed in  \cite{Zhitnitsky:2025bvy,Zhitnitsky:2026pai} in context of BL and  MT events, when UAP events   represent a  different stage of the same AQNs     propagating in atmosphere. In particular, it supports the observation on correlation  between MT events and UAP events as claimed   in \cite{Bruehl:2025}. It also supports the observation  of multiple MT events grouped together     as claimed  in \cite{Villarroel:2021ddh,Villarroel_2025,Villarroel:2025xym}.
 
  In the AQN framework this feature  (multiple MT objects)  and  multiple UAP correlated  objects  as reported in \cite{DOW-UAP-D108} is a natural consequence of the AQN framework as  both phenomena represent different stages   of the same AQN propagating in atmosphere. Many other features reported in  \cite{DOW-UAP-D108}, 
  }
 There are many features of numerous recently reported  events  (such as enormous velocity, high acceleration, the absence of the acoustic sound for supersonic motion, apparent presence of a faint trail, absence   of any remaining materials in affected area,  and many other puzzling features), as recorded  in   UAP reports  will be discussed in the present work within the same AQN proposal as these features are very generic of  this framework as mentioned in previous publications \cite{Budker:2020mqk,Zhitnitsky:2025bvy,Zhitnitsky:2026pai} and will be elaborated in the present work. Essentially, in this work we focus on specific class of UAP events  which could be explained by the AQN propagating in atmosphere, i.e, we identify: 
 \be
\label{identification}
{\rm [specific~ class ~ of~  UAP~events] } ~~\equiv ~~~ {\rm [AQN ~ induced~ events].}
\ee

In next Sect. \ref{AQN} we briefly overview the key  elements of the AQN framework relevant for the present  studies. The  central for this work  is  Sect.  \ref{differences} which 
contains a number of qualitative   estimates  demonstrating  how the puzzling and mysterious UAP features  could  be naturally explained within the AQN framework. 
% In   Sect. \ref{items} we briefly list the basic features of the UAP events along with recent   sixth UAP tranche    as reported in \cite{DOW-UAP-D108}.  
We argue that the list of puzzling features from Sect. \ref{items} finds a very natural explanation within AQN framework as we explain  in Sect.\ref{differences}.  The final Sect. \ref{conclusion}  represents our concluding comments along with    several   suggestions on   type of instruments required to  study this specific (and well defined) type  of the   UAP events representing  the  close cousins of BL,  MT and sky-quake  events in the AQN framework.
It  could provide an additional tools  for systematic studies of the UAP events   as recently advocated   in \cite{doi:10.1142/S2251171723400068} in context of the Galileo Project and in  \cite{Szydagis:2024eea} in context of UAPx project.

    \section{Brief overview of dark matter (DM) in the AQN framework}\label{AQN}
  Our goal here is to  briefly  explain  the basic ideas of the AQN framework. We want to avoid multiple repetitions of the same introduction on DM in form of the AQN  and refer to introductory section of ref.\cite{Zhitnitsky:2026pai} devoted to MT events. Here we give very few elements of the AQN construction  for completeness and self-consistency of the presentation.

The mainstream  idea for the last 40 years is based on    assumption  that   DM could be in form of some fundamental  Weakly Interacting Massive Particles (WIMP). However, the WIMP idea  has 
lost its initial   enthusiasm as dozen of dedicated instruments could not find any traces of WIMPs though the sensitivity of the instruments had dramatically improved by many orders of magnitude during the last decades. 

In the present work we consider a  fundamentally  different type of the DM which is in form of  
   macroscopically large composite objects of nuclear density, similar to the Witten's quark nuggets  \cite{Witten:1984rs,Farhi:1984qu,DeRujula:1984axn}.   The corresponding objects   are called the  AQNs and behave as    {\it chameleons}: they are (almost) 
    not interacting entities      in dilute environment, 
  such that the AQNs may serve as proper DM candidates.    However, the same objects interact very strongly with material when they hit the Earth, or other planets and stars. The AQNs are absolutely stable objects on cosmological scales as the energy per baryon charge in AQNs is smaller than the energy per baryon charge for nucleons.  
  
The main distinct feature of the  AQN model (which plays absolutely crucial role for  the present work) 
in comparison with old Witten's construction 
 is that AQNs  can be made of {\it matter} as well as {\it antimatter} during the QCD transition as a result of the charge segregation   process.      
   The presence of the antimatter  nuggets in the system implies that there will be  annihilation events leading  to very profound  strong effects when antimatter AQNs hit the Earth's atmosphere.  
More specifically, 
 our claim here is that the  basic features of specific type of the UAP  events as listed  below in  Sect. \ref{items}
   may  naturally emerge  as a result of these  powerful and energetic   events   related to antimatter nature of the AQN objects as described in Sect. \ref{differences}. 
   
   The AQN dark matter model was invented long ago  \cite{Zhitnitsky:2002qa} without any intension  to apply the model to UAP. Rather,   it was invented   with  a single motivation to explain the observed  similarity     $\Omega_{\rm DM}\sim \Omega_{\rm visible}$ between  visible and DM components, which  represents a very  generic feature of this framework. We refer to the original papers   \cite{Liang:2016tqc,Ge:2017ttc,Ge:2017idw,Ge:2019voa} devoted to the specific questions  related to the nugget's formation, generation of the baryon asymmetry, and  survival   pattern of the nuggets during the evolution in  early Universe with its unfriendly environment.
We also refer to    recent  original works  \cite{Majidi:2024mty,Sommer:2024iqp,Sekatchev:2025ixu,Majidi:2025ylh}      on applications of the AQN framework for  
different cosmological environments. 
 Finally,  we  refer to   
brief review \cite{Zhitnitsky:2021iwg} and    Les Houches  lectures  \cite{VanWaerbeke:2026sxs} covering all these topics with some details.

In the AQN scenario  the DM density, $\Omega_{\rm DM}$ representing the matter and anti-matter nuggets, and the visible    density, $\Omega_{\rm visible}$, will automatically   assume the  same order of magnitude densities  $\Omega_{\rm DM}\sim \Omega_{\rm visible}$   as they both proportional to one and the same fundamental dimensional parameter of the theory, the $\Lambda_{\rm QCD}$. Therefore, the AQN  model,   by construction, actually resolves two fundamental problems in cosmology  (explains the baryon asymmetry of the Universe, and the presence of   DM with proper density  $\Omega_{\rm DM}\sim \Omega_{\rm visible}$) without necessity  to fit any parameters of the model.  In other words,  the unobserved antibaryons in visible sector  in this model comprise 
dark matter being in the form of dense nuggets of antiquarks and gluons in  a specific type of quark matter.  Precisely these antimatter nuggets formed during the QCD transition in early Universe represent the source of energy 
for all the anomalous phenomena discussed in the present work as well as in previous papers   \cite{Budker:2020mqk,Zhitnitsky:2025bvy,Zhitnitsky:2026pai} devoted to studies of the  same objects  propagating in atmosphere, but manifested themselves in dramatically different ways.    
 
    The rareness of the AQN  events (which   according to (\ref{identification}) has  the same rate as UAP events) is related to tiny DM density in the Universe, i.e.
 \be
\label{Phi1}
\frac{  d \Phi}{  d A}
=\frac{\Phi}{4\pi R_\oplus^2}  =  4\cdot 10^{-2}\left(\frac{10^{25}}{\langle B\rangle}\right)\left(\frac{\rho_{\rm DM}}{0.3{\rm\,GeV\,cm^{-3}}}\right)
\left(\frac{v_{\rm AQN}}{220~ \rm km ~s^{-1}}\right)\rm \frac{events}{yr\cdot  km^2},
\ee
 where $R_\oplus=6371\,$km is the radius of the Earth, while $\langle B\rangle \approx 10^{25}$ is a typical baryon charge of the AQNs,   and  $\Phi$ is the total hit rate of AQNs on Earth:
\be
\label{Phi}
\Phi
\approx \frac{2\cdot 10^7}{\rm yr}  
 \left(\frac{\rho_{\rm DM}}{0.3{\rm\,GeV\,cm^{-3}}}\right)
\left(\frac{v_{\rm AQN}}{220~ \rm km ~s^{-1}}\right)
\left(\frac{10^{25}}{\langle B\rangle}\right), 
\ee
where $\rho_{\rm DM}$ and $v_{\rm AQN}$ are  typical  density   and     galactic velocity   of DM particles within Standard Halo Model (SHM).   
 
 In this work in our studies of the  UAP events we will use the same typical parameters of the AQN model which had been used  in all our previous estimates \cite{Budker:2020mqk,Zhitnitsky:2025bvy,Zhitnitsky:2026pai} related to anomalous phenomena on Earth.
 \be
 \label{eq:values}
   \la B \ra \approx 10^{25}, ~~~  \la M_{\rm AQN}\ra \approx m_p\la B\ra \approx 16 {\rm g}, ~~~ R\approx 2.25\cdot 10^{-5} {\rm cm}, ~~~ \rho_{\rm AQN}=\frac{\la M_{\rm AQN}\ra}{ \frac{4}{3}R^3}     \approx  3.5\cdot 10^{14}\rm \frac{g}{cm^3}. 
 \ee 
 %The basic parameters of this model (such as the typical baryon charge of the nuggets) 
% had been fixed also long ago by explaining the observed excess of radiation at different cosmological  scales.  In this work we use the same framework with the same set of parameters to study some    UAP events.   
  The same parameters had been also used
in explanation of  the observed excess of radiation  in different environments for dramatically different scales  \cite{VanWaerbeke:2026sxs}, including  well established and  very  puzzling  UV excess in our galaxy \cite{Sekatchev:2025ixu}.   
 
 There are two important points here: 1. Very small flux  (\ref{Phi1}) explains why AQNs had not been discovered yet in numerous conventional  DM detectors searching for WIMPs; 2.   The density  of  the object (\ref{eq:values})     is at least 13 orders of magnitude denser than  any  element on Earth made of  conventional {\it atomic material}\footnote{It is approximately the same  difference in scales between the average densities   of  a star  and neutron star (NS) with approximately the same masses, but dramatically different in sizes.   In fact one can think about nugget as a tiny NS   stabilized by  the axion domain wall  replacing  the gravity force in case of NS, see  reviews \cite{Zhitnitsky:2021iwg,VanWaerbeke:2026sxs}.}.   We want to illustrate   this key feature of the system by explicitly writing it down as follows:  
 \be
 \label{image}
\rm  [apparent~visible ~ size ~ of  ~``fireball" ] (~ meter~scale )~~ \gg ~~  [``solid" ~core~size ~of ~AQN ] (\mu m~scale ~R),
 \ee
 where the physical ``solid core" size of the AQN is determined by (\ref{eq:values}), while apparent (visible) size of AQN is determined by mean free path of the x rays in the atmosphere as we explain in next Sect.\ref{visible-size}. 
Precisely this property (\ref{image}) explains many puzzling features  related to the propagation of such object in atmosphere or in water,  
including well known and well recorded   phenomenon as passing BL through the glass window as explained in \cite{Zhitnitsky:2025bvy} within AQN framework. 
 
The property  (\ref{image}) is very different from everyday life  experience when visible and actual size of a classical  object are similar in values\footnote{ \label{misconception}This conventional assumption (on atomic nature of UAP  materials) leads to   a number of common misconceptions that any supersonic motion generates the boom in atmosphere and turbulence in water. This is not the case for the AQN-induced phenomena,  which precisely answers to the question formulated as the title of this work. Another misconception is related to an acceleration experienced by UAP as estimated, e.g. in \cite{e21100939} and required enormous power for such acceleration. We estimate the acceleration of the nuggets which could easily assume $\sim 10^3 g$ and more, without  contradictions to any laws of physics. The energy  which fuels  such puzzling dynamics  is entirely determined by tiny anti-matter nuclear AQN's core and measured in kW, similar to BL events,  see Sect.\ref{kinematics} with detail estimates for acceleration, power and relevant time scales.}. In contrast, 
 the visible apparent size of the AQNs in form of fireball  is determined by the emission features of the nuggets,  and measured in meters as will be explained in Sect.\ref{visible-size}, while physical ``solid" size of the object  itself
is measured in $\mu m$ according to (\ref{eq:values}). 

The  AQN with visible  apparent   size  actually represents   the plasma in atmosphere which is observed and recorded as the UAP. The energy which fuels the corresponding radiation from fireball  is originated from  much smaller (in size) nuclear  density   antimatter object with typical characteristics  (\ref{eq:values}). The total mass of the visible object is not measured in tons. Rather, it is measured in grams as it is determined by the air density surrounding the AQN.  From these comments it is clear  that the   UAP event as recorded in visible and IR in this framework is not made from a solid material, which is a conventional misconception as explained in footnote \ref{misconception}. Instead, it should be thought as  a gas of atmospheric plasma, emitting in $\gamma$ rays, X rays, visible and  IR bands, powered by  tiny antimatter material, similar to BL events as discussed in \cite{Zhitnitsky:2025bvy}.

For the purpose of this work we also need to overview the basic features of the interaction of the AQNs with surrounding material when AQN hits the Earth's atmosphere. We again refer to our previous papers 
\cite{Budker:2020mqk,Zhitnitsky:2025bvy,Zhitnitsky:2026pai} 
for the introduction into the subject in context of AQN propagating in atmosphere.  However, we need to list several important qualitative elements here to explain the observed UAP events within AQN framework.

i). When AQN enters the atmosphere the antiquarks  from AQN  start to annihilate 
 with surrounding material  which  leads to very high internal temperature of the AQN on the level of $T\approx$ 20 KeV. 
 This implies that the dominant emission from AQNs will be in from of X ray emission, in contrast with  conventional meteorites which normally assume the temperature in eV range,     such that they emit photons in optical frequency bands. 
 
ii). The high internal temperature $T$ of the AQN implies that a large number of weakly bound positrons  
from the electrosphere   can easily leave the system, similar to ionization of atoms at sufficiently high temperature.   As a result, the AQN   assumes  a negative  electric  charge $\sim -|e|Q$ with  $Q\sim 10^{11}$. The ratio $eQ/M\sim 10^{-14} e/m_p$ characterizing  this object  is very tiny. However, the charge $Q$ itself is sufficiently large   to  interact with  strong atmospheric electric field.

 iii). The propagating AQN in atmosphere with supersonic speed  will be mostly unnoticed as discussed in  \cite{Budker:2020mqk} because the dominant portion of the released energy  will be in form of the  X rays, axions and neutrinos, while the  energy released in the acoustic bands (mostly infrasound)  represents very tiny portion of the energy.   It should be contrasted with conventional meteorites (or other supersonic  objects  such as  jets) when acoustic boom   always accompanies   supersonic motion. This fundamental difference between high speed AQNs and any other conventional solid objects will be explained later in the text in  Sect.  \ref{differences}. On  microscopical level it is related to  the dramatic difference in the $E\&M$ spectrum (which consequently  generates acoustic shock wave) between AQNs and conventional  objects.    The difference in physical sizes as mentioned above (\ref{image})   also plays an important  role.   
 
 iv). A huge amount of energy being released as a result of propagation (in atmosphere, rocks, or water) does not imply that the energy will be released simultaneously as explosion\footnote{annihilation of $B\approx 10^{25}$ corresponds to 1.7 TNT tone equivalent if all energy released simultaneously in a single location.}. Rather, this energy will be distributed on fly along the entire trajectory   of the AQN as there is no enough material along the path to get annihilated simultaneously due to very tiny size of the object. Typical AQN with $B\approx 10^{25}$  with parameters (\ref{eq:values})  crosses entire Earth without loosing its mass and momentum, similar to Witten's strangelets. Due to the same reasons the AQN can propagate in water with enormous velocity because even in the dense water (in comparison to air)   there is no enough material along its path to  
 stop it. The source  which generates the required energy is the same antimatter material of   AQN which had been formed during the QCD epoch in early Universe. 
 %which cannot be released simultaneously 
 %(as in nuclear bomb), but rather it  is distributed on fly along its trajectory. 
      
v).   Occasionally, an external strong impact\footnote{it could be due to sudden increase of the annihilation rate within the AQN's quark core, which itself might be result of some local atmospheric perturbances such as  strong ionization  due to  thunderstorm or  as a  result of nuclear test \cite{Zhitnitsky:2026pai}. It may also a result of crossing   air-water interface with dramatic changes of the density along its path.}  may disintegrate AQN. In this case   a  small chunk from  the original AQN material (in form of the anti-baryon nuclear material) is separated from the original parent AQN, which can be viewed as a disintegration /separation /spallation  effect. 
%It will be obviously accompanied with enormous instantaneous energy release (explosion  like). From outside this event looks like a short lasting ``flash".
%For the present work precisely this ``flash" plays a key role as this moment of flash will be identified with MT event as discussed in Sect. \ref{sect:proposal}. 
% In a sense, it is very similar to well known and well studied    spallation effect which is very common phenomenon in nuclear physics. 
 The corresponding secondary particles have been coined in  \cite{Zhitnitsky:2025bvy}  as AQN$_s$, where subscript $``s"$ stands for  {\it  s}econdary particle or 
  {\it s}pallation.    
  The secondary AQN$_s$    have typical sizes $B_{AQN_s}\approx 10^{15}$ which is 10 orders of magnitude smaller than   original parent AQN with typical baryon charge $B\approx 10^{25}$, see (\ref{eq:values}). 
    %The corresponding fragment   really represents  very small chunk of the original (anti-matter) material, but in a different, not CS,  phase
 % in comparison with parent AQN.  The difference in binding energies   for parent AQN and daughter AQN$_s$ is the source of energy which should be released at the moment of spallation.   
 %These objects  should have typical sizes $B_{AQN_s}\approx 10^{15}$ which is 10 orders of magnitude smaller than   original parent AQN with typical baryon charge $B\approx 10^{25}$, see (\ref{eq:values}). 
 This typical value $B_{AQN_s}\approx 10^{15}$ was  extracted from  studies  of  BL  events within AQN framework \cite{Zhitnitsky:2025bvy}.  The same value  is found to be consistent with analysis of the MT events
  representing  the instant  of separation of AQN$_s$ from parental AQN. This event  of disintegration manifests itself as a short lasting ``flash". The same AQN$_s$ after separation will continue to propagate in atmosphere and manifest themselves as UAP  which represents the proposal (\ref{identification}).  These smaller chunks of antimatter, the daughter AQN$_s$ will also propagate and radiate  in visible and IR frequency bands along their paths on-fly,  rather than explode, see item iv) above.

   \section{AQN propagating in atmosphere: acoustic, visible and infrared signals }\label{differences}
   This section describes  the key elements of the proposal (\ref{identification}). It is heavily based on previous papers  \cite{Budker:2020mqk}, \cite{Zhitnitsky:2025bvy}, \cite{Zhitnitsky:2026pai} devoted to sky-quakes, Ball Lightnings (BL) and Mysterious Transients (MT),  representing themselves very  different manifestations   of the same AQN events. We describe expected radiation  from    AQN in $E\&M$ and acoustic  frequency bands.  We also discuss a number of kinematical features of the motion such as velocity and acceleration of the moving AQN bodies. We  emphasize on dramatic differences of all  these characteristics in comparison with  moving objects made of  conventional  materials such as meteoroids or supersonic jets. 
   
        \subsection{AQN in the Atmosphere. Acoustic signal }\label{acoustics}
   We start our review on   basic  characteristics of the acoustic  wave from propagating AQN by  estimating 
     the so-called  blast-wave relaxation radius defined as 
   \be
   \label{R_0}
  R_0^{\rm AQN}\equiv \sqrt{ \frac{E_l}{p_0}  },  ~~~ ~~~~ E_l \simeq   \kappa \cdot (\pi R^2) \cdot (2~ {\rm GeV})\cdot n_{\rm air} 
 \simeq 10^4\cdot \kappa  \left( \frac{B}{10^{25}}\right)^{2/3} \left(\frac{n_{\rm air}}{10^{21} ~{\rm cm^{-3}}}\right)\frac{\rm J}{\rm m}, 
   \ee
   where $E_l$ is the energy deposited by the moving body  per unit trail length  \cite{Budker:2020mqk}, and $p_0$ is the hydrostatic atmospheric pressure. 
   The physical meaning of this parameter $R_0$ is the distance at which the overpressure approximately equals the hydrostatic atmospheric pressure.  
 The parameter $\kappa$   in (\ref{R_0}) encodes a large number of complex processes including the probability that 
not all atoms and molecules may be able to penetrate into the 
quark phase of the nugget to get annihilated.  This parameter  also includes complicated dynamics due to the very large Mach number $M=v_{\rm AQN}/c_s$.  
It also includes the effective increase of the cross section in comparison wth geometric cross section $\pi R^2$ in ionized environment due to the Coulomb attraction. 
Numerically, the parameter  $R_0$ can be estimated as follows:
\be
\label{R_AQN} 
  R_0^{\rm AQN}\equiv \sqrt{ \frac{E_l}{p_0}  }\sim 0.3 ~ \left( \frac{B}{10^{25}}\right)^{\frac{1}{3}}\left({\frac{n_{\rm air}}{10^{21} ~{\rm cm^{-3}}}}\right)^{\frac{1}{2}}  {\rm m}.~~~~
\ee
 
 The significance of this parameter is as follows. 
    In the case of conventional 
   or nuclear explosions, the blast occurs as a result of the interaction of $E\&M$ radiation with surrounding material which rapidly heats the material. This causes vaporization of the material, in turn resulting in its rapid expansion, which eventually contributes to formation of the shock-wave. 
   All these effects occurring in conventional 
   explosions or supersonic motion 
     at spatial scales  much smaller than a typical radius where over-pressure approximately equals to atmospheric pressure.      
    
    The distances where the radiation is effectively converted to the
     shock-wave energy is determined by effective cross section. In the case of conventional, nuclear explosions or supersonic motion, e.g. meteoroids,  
    the dominant portion of   the radiation   normally occurs in the 1 to 20\,eV range, which includes visible light. At this energy, the dominant process is the atomic photo-effect with a cross section $\sigma_{p.e.}\sim 10^7\,{\rm ~barn} $ and higher, such that the photon   attenuation length  $\lambda \sim 10^{-6}{\rm g/cm^{2}}$, or even much higher. 
     These spectral features in air imply  that the energy due to heating is completely absorbed on spatial scales much shorter than $R_0$, i.e., 
   \be
   \label{lambda}
   \frac{\lambda}{\rho_{\rm air}} \lesssim 10^{-3}{\rm cm}\ll R_0 ~~~~~ [\rm   meteoroids]. ~~ 
   \ee
   This should be contrasted with the AQN case with a drastically different radiation spectrum with typical energy in the $\sim$20\,keV range as listed in item i) in Sect.\ref{AQN}.    Atomic photoelectric effect 
     is still the dominant process for this energy band
and the photon  attenuation length is $\lambda\sim  0.5\,{\rm g/cm^{2}}$ which is many orders of magnitude greater than for conventional explosions, i.e.  
     \be
   \label{lambda_AQN}
   L=\frac{\lambda}{\rho_{\rm air}} \simeq 5 ~{\rm m}\gg R_0^{\rm AQN} ~~~~~ [\rm AQN~ events]. ~~ 
   \ee
   These estimates imply two key elements related to acoustic signals: 1.    only a small portion of the energy (\ref{R_0}) will be released in the form of a blast, while the dominant portion of the  energy will heat the surrounding material; 2. the scale where conversion of $E\&M$ energy to acoustic energy occurs is dramatically different for two cases: $10^{-3}$ cm (\ref{lambda}) versus 5m for AQN (\ref{lambda_AQN}).  This large attenuation scale for AQN leads to vaporization and heating without rapid expansion. This effect  leads 
   to a weak acoustic signal in comparison with conventional explosions and supersonic motion, e.g. meteoroids.

   Indeed, the overpressure from AQN events can be estimated as  \cite{Budker:2020mqk}
 \be
\label{AirDeltaPnumber}
 \delta p\approx 0.03\,{\rm Pa} \left( \frac{B}{10^{25}}\right)^{2/3}\left( \frac{100\, {\rm km}}{r}\right)^{3/4}~[\rm AQN~ events].~~~~~
 \ee
 This estimate  shows that  a typical AQN generates a small overpressure $\delta p/p \sim 10^{-3}  $ even inside the absorption region $r\lesssim L$, which  should be contrasted with the meteoroid case   where $\delta p/p \approx   1$
 at $r\simeq R_0$. The difference as we already mentioned is due to the large   length $L\simeq 5$\,m in comparison with the small absorption distance $\sim 10^{-3}$ cm  for the  meteoroids (\ref{lambda}).

      Another important   characteristic of the acoustic  waves produced by AQN is very low frequency band in comparison with conventional explosions or supersonic motion, e.g. meteoroids.    In the AQN  case we arrive at the following estimate for the frequency at a distance $r$:
\be
\label{frequencyAQN}
\nu (r) 
 \sim 6\, {\rm Hz} \left( \frac{100\, {\rm km}}{r}\right)^{1/4}  ~[\rm AQN~ events].~~
\ee
  We conclude this subsection with the following remarks:
  
  1.  While estimate for the overpressure is very tiny according to (\ref{AirDeltaPnumber}) the signal from moving AQN still can me recorded with existing instruments for sufficiently large nuggets with $B\gtrsim 10^{27}$. In fact,  a  rare mysterious event  took place on July 31, 2008, which was properly recorded by the Elginfield infrasound array (ELFO) near London, Ontario, Canada \cite{ELFO}. It  was identified as  AQN event in Ref. \cite{Budker:2020mqk}, and it was argued that the energetic properties (\ref{AirDeltaPnumber}), infrasound frequency characteristics (\ref{frequencyAQN}), and other attributes of the AQN event are consistent with the observed ELFO phenomenon.

  2. The dramatic difference in photon  attenuation lengths between  conventional explosions, or  supersonic motion (\ref{lambda})  versus the AQN  supersonic motion  (\ref{lambda_AQN}) quantitatively answers to the question formulated in  the title of this work. To be concrete:  the  absence of a strong  acoustic boom for mysterious UAP events,  within AQN framework (\ref{identification}), is a result of very hard spectrum of the annihilation events (in keV range) in contrast with typical spectrum  for any other conventional explosions  and supersonic motion (in eV range).   The dramatic difference in photon's   attenuation lengths for keV and eV photons 
  explains the absence of the boom for supersonic motion within AQN proposal (\ref{identification}). This important  feature is explicitly listed as item iii) in Sect \ref{AQN}.
  
  3. The collisions of the AQN with surrounding material  does not affect   the AQN trajectory  as there is no enough material along its path to modify the momentum due to the enormous nuclear density of the AQN core (\ref{eq:values}). It is very similar to Witten's nugget which is   capable  to cross the Earth without changing its velocity, see item iv) in Sect \ref{AQN}.
  
  The situation is dramatically different in case when small chunks of matter    get separated from the parent AQN to form much smaller secondary AQN$_s$ as  mentioned  in item v)  in Sect \ref{AQN}. The secondary AQN$_s$ due to their much smaller masses and much larger ratio $(R^2/M_{\rm AQN})$  can effectively slow down  in the atmosphere with enormous magnitude for acceleration exceeding $10^2$g and even $10^3$ g. The secondary AQN$_s$  can stop and even change the direction of motion as discussed in  \cite{Zhitnitsky:2025bvy} in context of the BL events. The same features  should  also apply to UAP events if proposal (\ref{identification}) is adopted, as will be discussed in next subsections \ref{kinematics} and \ref{E-field}. 
  
  \subsection{AQN$_s$ (apparent) size in visible  and IR  frequency bands}\label{visible-size}
  In this and following subsections we discuss secondary AQN$_s$ after spallation with much smaller $B_s\sim 10^{15}$ in comparison with original AQN. 
The corresponding estimates have been carried out in \cite{Zhitnitsky:2025bvy} in context of the BL events, which are close cousins of UAP events  in AQN framework. We overview the corresponding estimates below. The UAP events likely to have larger size and baryon charge in comparison with BL events by few orders of magnitude. However, we opted to keep our original estimates  \cite{Zhitnitsky:2025bvy} relevant for BL events below as the rescaling to larger AQN$_s$ can be easily performed when  needed, see next Sect.\ref{kinematics} with corresponding scaling factors. There are few elements which are important for understanding   the kinematics and dynamics of  the AQN$_s$ which are qualitatively different from previous Sect. \ref{acoustics} devoted to studies of much larger objects. First of all the internal temperature of the  AQN$_s$ is few times smaller than 20 keV which was the benchmark for AQN as discussed in item i) in Sect \ref{AQN}.  For the  AQN$_s$ it was estimated on the level of $T\approx 6$ keV mainly due to much smaller  velocity of the propagating object as the scattering processes (friction) are   much more efficient for smaller  AQN$_s$ in comparison with parent AQN. Precisely this temperature  $T\approx 6$ keV determines the apparent (visible) size of the AQN$_s$.

This paragraph explains the nature of the apparent (visible) size of the $AQN_s$. There is a common  confusion between  apparent (visible) size of the $AQN_s$ and actual, physical size of the $AQN_s$ nuclear core (\ref{eq:values}). This confusion  leads to a   misconception on expected behaviour of such objects,  see footnote \ref{misconception}. 
  The key feature of the AQN radiation   is a broad spectrum with   $\omega\lesssim  T$. What happens to these keV photons emitted by  the  $  AQN_s$  
    when internal   core temperature is around $T\approx 6$ keV ? First, these photons get absorbed. 
   The dominant mechanism of absorption for such energetic X rays is the atomic photoelectric effect with cross section $\sigma_{p.e.}\propto \omega^{-7/2}$.
     Based on this cross section the mean free path $\lambda $ for atmosphere at sea level for 6 keV photons can be estimated as \cite{Zhitnitsky:2025bvy}:
    \be
   \label{lambda1}
   \lambda^{O,N} {\rm (6 ~keV)\approx 10 ~  cm}, ~~~  \lambda^{O,N}\propto T^{\frac{7}{2}}\exp\left(\frac{h}{8~ \rm km}\right) ~~~~~ [\rm apparent~visible ~size~ of~  AQN_s].
   \ee
  % However, dependence of $\sigma_{p.e.}$ on frequency of emitted photons is very strong $\propto \omega^{-7/2}$, and absorption length could easily reach few meters at higher altitudes for higher temperature. 
    The atomic photoelectric effect is accompanied by emitting electron   ionizing atoms and molecules along its path. The life time of a free electron in air  is  very short, around $0.1\mu \rm s$ such that free electron will be quickly absorbed by atoms in air on very short distances around $10^{-4} \rm m$, much shorter than (\ref{lambda1}). Furthermore, the $AQN_s$ ionizes surrounding material along its path.  These excited and ionized atoms and molecules   will emit visible and infrared light and will heat the surrounding material. 
    This volume of air with typical scale (\ref{lambda1}) will be observed from a distance   as a radiation coming from UAP according to our proposal (\ref{identification}). This region can viewed as the plasma of ionized gas emitting   the photons in broad frequency bands. 
    Therefore, we identify the  apparent  size of UAP (in visible and IR frequency bands) with mean free path $\lambda$ in surrounding air as stated in (\ref{lambda1}).

    This scale (\ref{lambda1}) derived for the atmosphere, and its generalization for solid materials or water,  is perfectly consistent with observations  in context of the BL physics (including well reported phenomenon of BL passing  through a  window glass)
    as directly  measured by optical and scanning  microscopes and laser beam probing the glass, see  \cite{Zhitnitsky:2025bvy} for references and details.  
    
    As we argue below in next Sect. \ref{items} some of the UAP events could be also related to the same AQN induced events with apparent (visible) size represented by hot plasma  surrounding the core of the $AQN_s$. This hot gas of ionized air must emit in IR and produce vapour as it cools down. The size of the object in IR must be the same order of magnitude as the volume of the hot gas where 
 $AQN_s$ deposited its energy. The apparent size of the object (\ref{lambda1}) is obviously very different from nuclear core scale (\ref{eq:values}), which is explicitly illustrated by        (\ref{image}).  The shape of the flying object should  typically assume   a symmetric spherical form. However due to a high  velocity of the $AQN_s$ and atmospheric air motion the observable shape could assume a prolongate geometry  as the observed shape  represents the atmospheric ionized plasma, not the sold nuclear core material as mentioned above.  
 
 As we already mentioned this picture of emission  is consistent with numerous observations in the BL context \cite{Zhitnitsky:2025bvy}.
In particular,  it is known that the  BL emits UV,  x rays and $\gamma$  radiation as it has been directly   observed and recorded \cite{STEPHAN201632}. This hard emission 
  must be present in the system as it is originated from the  core of the $AQN_s$, in contrast with visible and IR emission  
   which is a secondary process in the AQN framework as described above.     This hard emission is unique feature of the proposal (\ref{identification}) as no any other flying objects (e.g. meteorites or drons) can emit hard radiation. 
   
  Another remark  is that the apparent  size of the BL in this proposal (\ref{identification}), being identified with $\lambda$,  is not related to a basic energetic characteristic  of the UAP events (the baryon charge $B_s\gtrsim 10^{15}$), which could dramatically  vary from case to case.     Rather, the variation in size of the UAP (in visible and IR frequency bands) is related to the  internal temperature $T$ of the objects and atmospheric conditions which ionizes the plasma surrounding the nuclear core. The intensity of the radiation, on other hand, is highly sensitive to baryon charge $B_s$ of the $AQN_s$ and could vary by many orders of magnitude as we discuss in next subsection \ref{kinematics}.

\subsection{kinematical characteristics:  velocity,  acceleration, life time, power of radiation }\label{kinematics}
The typical effective size  of the  AQN$_s$ is   estimated from the   condition that the binding energy of the positrons in electrosphere is approximately equal to internal temperature $T\approx 6$ keV of the system \cite{Zhitnitsky:2025bvy}:
 \be
 \label{R_eff}
   \frac{\alpha Q_{\rm AQN_s}}{R_{\rm eff}}\approx T\approx 6 ~ {\rm keV}~ \rightarrow ~R_{\rm eff}\approx 10^{-6} \rm cm  ~~ (for ~Q_{\rm AQN_s}\approx 6\cdot 10^4 ~~ and ~B_s\approx 10^{15}).
   \ee
 The scale $R_{\rm eff}\approx 10^{-6} \rm cm$ should be   identified with internal effective  size of an $AQN_s$ as the bound positrons will always stay in vicinity of  the nuclear matter core due to the strong electromagnetic attraction. It should not be confused with visible size  (\ref{lambda1}) which is mostly filled with atmospheric  gas. Precisely this scale $R_{\rm eff}$ determines the  required 
  total amount of material which may stop the nugget. In other words,   the total amount of the material in  a cylinder of  radius $R_{\rm eff}$ and length $L$ must be the same order of magnitude  as the baryon charge $B_s\approx 10^{15}$  for the $AQN_s$ to   loose its huge initial velocity, i.e.
     \be
\label{eq:stop}
  B_s\sim    (\pi R_{\rm eff}^2) \cdot  n_{\rm air} \cdot L~~~~~\rightarrow ~~~~  L\sim 3~\rm km,
\ee 
where     $n_{\rm air}\approx 10^{21}\rm cm^{-3} $ is the baryon density of the surrounding atmospheric gas. The length $L$ has a physical meaning of a distance where the $AQN_s$ can dramatically slow down from its initial velocity $\sim 10^{-3}c$ at the moment of spallation. 

One can also estimate the stoppage time $\tau_s\sim 2$s before $AQN_s$ reaches relatively low velocities $v_{\rm air}\sim 0.5 \rm km/s$, close to the speed of sound,  representing  a typical velocity  of air molecules as estimated   in \cite{Zhitnitsky:2025bvy}. Similarly, one can also estimate the life time $\tau$  of the $AQN_s$ when the fuel in form of the antimatter nuclear material is completely consumed as a result of steady annihilation processes when $AQN_s$ propagates in atmosphere. The corresponding estimates had been performed   in context of the BL studies for $B_{\rm AQN_s}\approx 10^{15}$, and we generalize the corresponding    estimate for larger baryon charges relevant  to UAP events:
\be
\label{eq:tau}
\tau\approx 10 ~{\rm s}  \left( \frac{B_{\rm AQN_s}}{10^{15}}\right)^{1/3}\cdot \left( \frac{10^{21}\rm cm^{-3}}{n_{\rm air}}\right),
\ee
where scaling   $\propto B^{1/3}$ in this formula is due to the fact that the rate of the annihilation processes powering the UAP is proportional  to the surface of the nuggets, i.e. $R^2\propto B^{2/3}$, while the total amount of the antimatter material to be annihilated during the UAP's life time is proportional to the volume, i.e. $R^3\propto B$.    

The total power of radiation $P_{\rm tot}$, inclduding all frequency bands,   can be estimated in terms of life time $\tau$ as follows 
 \be
      \label{eq:power}
      P_{\rm tot}\approx \frac{B_{\rm AQN_s}\cdot m_p}{\tau}\approx   \frac{10^2 {\rm kJ}}{\tau} \cdot \left( \frac{B_{\rm AQN_s}}{10^{15}}\right)^{2/3} \approx 10 ~{\rm kW} \cdot \left( \frac{B_{\rm AQN_s}}{10^{15}}\right)^{2/3},
       \ee
where benchmark value $10 ~{\rm kW}$ was estimated in \cite{Zhitnitsky:2025bvy} in context of the BL events for $B_{\rm AQN_s}\approx 10^{15}$.  On should emphasize that the radiation in visible and IR bands represents a small fraction (around few percent) of the total power of radiation (\ref{eq:power}) as the dominant fraction of radiation will be deposited into the surrounding air in form of the heat, UV and hard radiation, see item 4 below.

One can also estimate the acceleration (in fact deceleration)  of such objects when they experience strong friction due to collisions and annihilation events with surrounding material after spallation time at instant $t_0$, see \cite{Zhitnitsky:2025bvy}:
 \be
 \label{eq:acceleration} 
a_{v_{\rm AQN_s}} \approx 1.5\cdot 10^3 {\rm \frac{m}{s^2}}\cdot \left(\frac{\tau_s}{(t-t_0)}\right)^2\gg 10^2 g.
 \ee
  In fact it could easily exceed $10^4$g at $(t-t_0)\approx 0.1 \tau_s\approx 0.2 s$ when the velocity of the $AQN_s$ already dropped by 90\% from  its original value to reach $30~ \rm km/s$. Such enormous acceleration 
 had been mentioned previously in the literature \cite{e21100939} in relation with UAP event.

 We conclude this subsection with the few remarks:
 
 1.  The $AQN_s$ start the propagation in atmosphere (after separation from parent AQN) with enormous velocity $\sim 10^{-3}c$ without producing any observable acoustic booms as overpressure (\ref{AirDeltaPnumber})  is tiny for $B_s\approx 10^{15}$.  After that it efficiently slows down with a typical stoppage time $\tau_s$ measured in seconds. 
 % is estimated from condition that   $AQN_s$ looses its 99\% of its momentum when it reaches $\rm v_{AQN_s}\sim km/s$ range.
The $AQN_s$ looses its 99\% of its initial velocity  when it reaches $\rm v_{AQN_s}\sim km/s$ range,  after which  the object can be observed  
 in visible and IR bands because the internal temperature $T$ is dropped, see Sect. \ref{visible-size} for estimates. The same effect obviously occurs for BL events 
 when they become visible close to the ground after travelling $L\sim 3 \rm km$ or so from thunderclouds\cite{Zhitnitsky:2025bvy}. 
  
2.   The same comment obviously applies to the $AQN_s$ hitting the water and consequently propagating  in water with the same initial speed. However, the typical value for length $L$ in water  is dramatically shorter
(for the same value of $B_s$) in comparison with propagation in atmosphere due to much larger density of the surrounding media with $n_{\rm water}\sim 10^{24} \rm cm^{-3}$ replacing 
$n_{\rm air}$ in eq.(\ref{eq:stop}). It becomes larger for larger $B_s$ according  to eq.(\ref{eq:tau}) when $n_{\rm air}$  should be replaced to $n_{\rm water}$.  There will be  no turbulence nor any other perturbations related to this motion as these $\mu m$ size objects are too small to generate any  shock waves. The apparent visible size of the objects will be much larger in size as stated  in (\ref{image}). Precisely this feature answers the question formulated as the title of this work. 
%The corresponding estimates will be presented    in Sect. \ref{visible-size} with details. 
%We emphasize that  the visible size    of the object   does not represent the actual ``solid" scale  of the $AQN_s$, similar to our discussion \cite{Zhitnitsky:2025bvy} in context of the BL physics.

 3.  The same comment also  applies to the $AQN_s$  initially propagating in atmosphere  and consequently hitting  a rock. After  the hit  it continues to propagate under the surface. This happens when  $B_s\gg 10^{15}$ is relatively  large and the amount of the atmospheric material along the trajectory is not sufficient to stop the object.  In this case one cannot see a propagating $AQN_s$ in deep underground, in contrast with propagation  in water mentioned above.  However, the location of impact should leave some traces in form of hot and melted substance  with identically the same chemical composition as surrounding material\footnote{In fact, there are many recorded     events, which are classified as ``Pseudo-meteorites  events" when a meteor-like event is observed, but no actual physical meteorite  is found in the area. We identify these ``Pseudo-meteorites  events" with AQN-induced events with $B_s\gg 10^{15}$.}.

 4. The AQN-induced events should   also leave some traces  of hard radiation  in affected area. 
 This is because  the hard radiation is inevitable feature of the AQN induced processes.  In fact, it has been  reported that some BL  witnesses who had close contacts with BL events had  developed some symptoms of radiation. Further to the same point: the fact that BL emits UV and   x ray    radiation   has been directly   observed and recorded as discussed in \cite{STEPHAN201632}.

\subsection{Motion in atmospheric electric field  }\label{E-field}
In previous subsection we discuss the kinematics of the $AQN_s$ due to the collisions and annihilation processes with surrounding material. It leads to  enormous 
deceleration (\ref{eq:acceleration}) which may efficiently slow down the object on  very short time scales $\tau_s$ measured in seconds. 

There is another external factor which may dramatically impact the behaviour of the flying objects, it is presence of the  electric field in atmosphere. It is known that the electric field could be enormous in atmosphere especially during thunderstorms.  For example, the  pre-existing electric field ${\cal{E}}$ plays a crucial role in dynamics of the  lightning processes. This field must be correlated on enormous scale  $L_{\cal{E}}$, which   essentially determines  the allowed scale of the   inhomogeneity  and   non-uniformity of a fluctuating   electric field for  the exponential growth  to hold for sufficiently long time (the avalanche). The  pre-existing electric field ${\cal{E}}$ and its correlation length $L_{\cal{E}}$ could easily reach the following values \cite{DWYER2014147} \cite{Gurevich_2001}:
\be
\label{E}
{\cal{E}}\gtrsim  \rm  \frac{kV}{cm},   ~~~ L_{\cal{E}}\gg (50-100)~ {\rm m}.
\ee

Now we want to address the following question: can the trajectories of  a massive   classical objects with typical parameters (\ref{R_eff}) in form of the $AQN_s$ be impacted by such atmospheric fields? The ultimate answer is ``yes" as estimates below suggest.  Indeed, one can estimate a typical  acceleration $a_{\cal{E}}$  experienced by the $AQN_s$ in the background external field (\ref{E}) assuming our benchmark parameters (\ref{R_eff}) extracted in context of BL studies:
\be
\label{a}
a_{\cal{E}} \approx \left(\frac{e {\cal{E}}}{m_p}\right)\cdot \left(\frac{Q_{\rm AQN_s}}{B_s}\right)\approx 0.6\cdot 10^3
\rm \frac{m}{s^2} \cdot \left( \frac{\cal{E}}{\rm kV/cm}\right)\cdot  \left( \frac{10^{15}}{B_{\rm AQN_s}}\right)^{1/3},
\ee
which is the same  order of magnitude  as  the acceleration (\ref{eq:acceleration}) due to the  friction forces. However, there is an  important distinction between the two as  atmospheric electric field fluctuates in space and time.  As result,  it may dramatically change the trajectory of $AQN_s$ if the object already slows down to the sufficiently  low values. In what follows for our numerical estimates we take  the benchmark  for ``low" velocity the value $v_{\rm air}\sim 0.5 \rm km/s$   as explained after (\ref{eq:stop}).

Our next task is to estimate a typical time $t_{\cal{E}}$ and distance $l_{\cal{E}}$ when the $AQN_s$ may change its direction as a result of  atmospheric electric field:
\be
\label{l}
t_{\cal{E}}= \frac{v}{a_{\cal{E}}}\approx 0.5 {\rm s}  \left( \frac{\rm kV/cm}{\cal{E}}\right)\cdot \left(\frac{v}{0.5 \rm km/s}\right)\cdot\left(\frac{B_{\rm AQN_s}}{10^{15}}\right)^{1/3} , ~~~~~~~     l_{\cal{E}} =\frac{v^2}{2a_{\cal{E}}}\approx 2\cdot 10^2 {\rm m} \left(\frac{v}{0.5 \rm km/s}\right)^2 \cdot\left( \frac{\rm kV/cm}{\cal{E}}\right). 
\ee
The typical scale $l_{\cal{E}} $ as estimated above  is the same order of magnitude as  the  correlation length $L_{\cal{E}}$ for  fluctuating electric field  in atmosphere as given by  (\ref {E}). Therefore, our estimate (\ref{l}) suggests that relatively slow (still close to the speed of sound) $AQN_s$ can indeed change directions due to atmospheric electric field as such field with such correlation lengths   could  be present in atmosphere at the moment when $AQN_s$  travelling  in atmosphere.

  %\subsection{AQN$_s$ in   infrared frequency bands}\label{visible-size}
  %As we mentioned previously, most of the energy released by AQN$_s$ is deposited to space in form of the heat  \cite{Zhitnitsky:2025bvy}, and must be observed in IR.  

 $\bullet$We conclude this section with the following comment. 
There are no any inconsistencies with law of physics when a classical  object can move with supersonic  velocity, shows  enormous acceleration (\ref{eq:acceleration}), and does not produce  any sound as  argued in this section. This anomalous behaviour  could not be explained with a common assumption that the objects are made from conventional {\it atomic material}. Precisely this assumption leads to a number of misconceptions as remarked in footnote  \ref{misconception}. Rather, the AQN is made of {\it  nuclear}, much denser, material  (\ref{eq:values}). These  objects  have apparent macroscopical size  (\ref{lambda1}) which can be observed in visible and infrared as a plasma fireball surrounding the nuclear core of the AQN$_s$. This apparent size     is dramatically different from their actual physical size as eq.  (\ref{image}) states. One should emphasize, that a similar concept on nature of BL events (which represent the UAP  mini cousins in the AQN framework) recorded for centuries can explain simultaneously {\it all} mysterious and puzzling manifestations  within the same AQN proposal as argued in  \cite{Zhitnitsky:2025bvy}. We also refer to the same paper as it contains large  number of conceptual questions (and answers) on  AQN framework from two anonymous Referees (25 questions from Referee-1 plus 34 question from Referee-2) which can be openly viewed by the readers. We also refer to \cite{QCD-cosmology} as illustrative  simplified introduction into an unorthodox  DM paradigm when the DM objects are  composite  nuggets made of   quarks or antiquarks   of the Standard Model (SM) physics. 
Hopefully it answers many conceptual questions  which may occur  while reading the present work.

   \section{Brief overview of puzzles and mysteries  of    UAP events from AQN perspective }\label{items}
 There are many different types of UAP events which manifest themselves in  dramatically different ways. It is obviously not the goal of the present work to overview 
 and classify these events. Our goal is dramatically different: we want to describe very small sample of UAP events which had been recorded  and which   show some similarities  with generic features of AQN described in previous Sect. \ref{differences}.  For completeness of the presentation we overview the puzzling UAP features below and 
   refer to a specific equations or sections where similar properties had been discussed within AQN framework  in Sect. \ref{differences}.

 %\subsection{Basic features of   UAP events   within AQN framework}\label{UAP_events}
  
  \subsection{``Green Fireballs and Project Twinkle"}\label{green_fireball}
  We start with very famous and historically very important project   ``Green Fireballs and Project Twinkle"  as reviewed in  \cite{knuth2025newscienceunidentifiedaerospaceundersea}. 
  %We quote few lines from  Section 3.3.5  of this paper. 
   %  with description of observations  which closely resemble the manifestations    of the AQN$_s$ events as given in previous Sect.\ref{theory}.      
      In late November of 1948 a   number of green fireballs had been reported, see large number of references  in  \cite{knuth2025newscienceunidentifiedaerospaceundersea}.
 The investigation started when  Kirtland Intelligence Officers contacted astronomer and meteor pioneer, Lincoln La Paz
from the University of New Mexico. Now we quote few paragraphs from Section 3.3.5 of \cite{knuth2025newscienceunidentifiedaerospaceundersea} to give a flavour of the observed events given by professionals: ``La Paz insisted that the
green fireballs could not be meteors because their trajectories were too flat,
their colour too green, and that no meteoritic material had been recovered".  Another quote is: ``But the biggest mystery of all was the fact that no particles of a
green fireball had ever been found. If they were meteorites, La
Paz was positive that he would have found one."  Another unusual factor was a large size of the green fireballs which was described as ``terrifying".  An enormous observed velocity 
of the  the green fireballs, which is not typical for the meteorites, was also emphasized by eyewitnesses. 
Such large meteorites moving with enormous velocity normally are accompanied by sound and shock waves that
break windows. Yet,  the observers reported that they did
not hear any sound.
One additional quote:
``if the green fireballs were meteors, then they should have been observed
globally --certainly not limited to a handful of sensitive nuclear sites in New
Mexico". Our final comment is that the flying objects may  generate  strong signal in hard radiation bands which was observed by Harley Marshall at the  Naval Electronic Laboratory  
Geiger counter, see Section 3.3.5 of \cite{knuth2025newscienceunidentifiedaerospaceundersea}.  

Now we are in position to address all these mysterious elements in behaviour of fireballs: 

1. from AQN viewpoint any remaining  meteoritic material  could not be recovered because the annihilation   of the AQN  with surrounding material leave the heat and radiation without any traces of ``particles from meteoritic material" as it was not a meteorite; 

2. the size of the plasma fireball could be very large  
according to (\ref{lambda1}) as it is determined by the internal temperature $T$ and not by the amount of material which could   fill   enormous volume $\sim \lambda^{3}$ determined by the mean free path of the X rays. The density of ``material" of  this large volume   is the density of the air surrounding the antimatter  nuclear AQN's core. This apparent size is  not related to   any ``solid material" which  represents   a common misconception in interpretation of the UAP events, see footnote \ref{misconception}. 

3.  enormous velocity of the object is also a common feature of the moving AQN as they enter the atmosphere with velocity $v\sim 10^{-3}c$ and  slow down to $v_{\rm air}\approx \rm 0.5  km/s$ as discussed in \ref{kinematics};

4. any acoustic waves cannot be heard  from  a  typical AQN event because it produces  very weak signal, and  it  mostly generates infrasound waves, as discussed in \ref{acoustics}.
A specifically designed  instruments are required to record such sound waves as we discuss below in context of ELFO event which took place on July 31,2008 in Canada. The anticipated boom from  flying objects of ``terrifying" size   with enormous mass  is related to a common misconception as summarized at the end of Sect. \ref{differences} and also  in footnote \ref{misconception}. The mass of these objects is in fact very tiny as it is determined by the air density of the surrounding gas as explained in item 2 above;

5. A non-observation of such fireballs globally also finds its natural explanation within AQN framework. Indeed, the  events had been observed in sensitive nuclear sites in New Mexico in 1948 where nuclear tests had been conducted on site. From AQN perspective the presence of the UAP events is  the consequence of a strong ionization of the atmosphere in the area as a result of these nuclear tests. Indeed, the key element of the proposal (\ref{identification}) is spallation mechnism, when smaller pieces are produced as stated in item v)  in Sect. \ref{AQN}. Precisely these daughter AQN$_s$ manifest themselves as BL events \cite{Zhitnitsky:2025bvy} or MT events  \cite{Zhitnitsky:2026pai}. In context of the BL events the ionization of the atmosphere was present in the system as  BL events are strongly correlated with thunderstorms where such strong ionization is known to occur. In context of the MT events the correlation with nuclear test had been also recorded. Therefore, it is not really a surprise from AQN perspective that UAP events had been observed in sensitive nuclear sites in New Mexico in 1948, and not globally where ionization was much smaller. 

6. The hard radiation is inevitable feature of the AQN proposal  (\ref{identification}) due to  x rays,  $\gamma$ rays and   strong ionization of the air which always   accompanies   the propagating   AQN.  At the same time the  emission in visible and IR is due to the secondary processes.  Therefore, it is not really a surprise from AQN perspective that Geiger counter had recorded the signal correlated with flying objects. The hard radiation has been also directly recorded in context of BL physics, see  item 4 in Sect.  \ref{kinematics}. 

 \subsection{``Pseudo-meteorites  events"}\label{sect:ELMA}
 
           There are large number of reports in the literature describing the meteor-like events which  (after detail studies by professionals) turned out to be not the meteorites. Such type of  events are often classified as ``Pseudo-meteorites  events", see   \cite{2020arXiv201200686O} with large number of  references. In what follows we  focus on a one specific  event   which reflects the basic feature of the UAP events, when a  fireball is observed but no any materials are found on the site. The event  took place in US town of Elma (Washington State) on July 15, 2003 as described in \cite{2020arXiv201200686O}.  
          
     The key elements of the Elma event are \cite{2020arXiv201200686O}: 1.after the event (fireball) the   glassy black rocks had been found; 2. the  rocks were very hot (one of the witnesses had burned a thumb and a finger by collecting the rocks); 3. analysis of the discovered stones by the University of Washington concluded  that the glassy rocks were not meteorites, see pictures  in  \cite{2020arXiv201200686O}. 
     
 From AQN perspective this event is very similar to the ``green fireball" events described above,   as no particle material is found on site. 
As we discussed  at the end of  Sect. \ref{kinematics}  in item 3  the $AQN_s$ with   relatively  large $B_s\gg 10^{15}$
     cannot efficiently slowdown within atmosphere, and it   is likely to hit the earth's surface with a high speed. As a result,  the $AQN_s$  gets annihilated in deep underground regions  by heating the surrounding rocks along its path. In many respects the heating of the surrounding material is similar to phenomena of the  BL passing through the glass window when the internal $AQN_s$ temperatures could reach enormous values and melt the glass and pass through it \cite{Zhitnitsky:2025bvy}.   It explains the hot glassy   rocks found in the area with chemical composition identically the same as in surrounding area. 

      \subsection{ELFO event as sky-quake event}\label{sect:ELFO}

      An  AQN enters the Earth's atmosphere with a typical for DM velocity $v_{\rm AQN}\sim 10^{-3} c$. In majority of cases (as discussed in item iv at the end of  Sect.\ref{AQN}) it propagates through atmosphere and Earth's core without loosing much of momentum and baryon charge
     as the cross section of the interaction  is relatively small, and  can be well approximated by its geometrical size $\pi R^2$, where $R\approx 2.25\cdot 10^{-5} \rm cm$ is the AQN's core size   for a typical $B\approx 10^{25}$ according to (\ref{eq:values}). 
     Such events can be hardly noticed by eyewitnesses as this  object does not emit  much  in  visible frequency bands. Instead, it mostly radiates  the X rays \cite{Budker:2020mqk}.    
     Nevertheless, in rare cases when the AQN  is sufficiently large in size with $B\approx 10^{27}$ and larger such  events can be observed due to their relatively strong shock  waves which can be recorded by dedicated  infrasound instruments such as Elginfield Infrasound Array (ELFO), see below.

      We have to make a short detour here on history of similar  puzzling very rare events, the so-called skyquakes.    The skyquakes,    mysterious   events which  have been recorded for centuries, similar to BL events. However, they  cannot   be explained by conventional physics despite extensive efforts \cite{skyquakes}. We refer to an interview \cite{skyquakes-meteorologist} as given  by a meteorologist to TV  station for a short introduction into a subject. The main message from this brief interview is that skyquakes cannot be attributed to seismic activities nor meteor-related events, which are routinely recorded around the world.  We also want to mention  another  powerful and rare event that occurred in Alabama on a clear day in 2017, when a sound was heard across 15 counties \cite{NASA}. There were no meteor-activity reports associated with this event; no visible lights, which typically accompany meteoroids, were recorded. However, there were reports of vapour trails.  As we already mentioned in discussions after formula  (\ref{lambda_AQN})  the presence of vapour is expected and even predicted by the proposal \cite{Budker:2020mqk} that associates skyquakes with AQN-induced infrasound acoustic shock waves\footnote{One may wonder why   very large AQN with $B\approx 10^{27}$ had not been noticed by all  sky camera  \cite{Budker:2020mqk}, while the events with much smaller  $B\gtrsim 10^{15}$ such as BL events or `` green fireballs" had been observed in visible frequency bands. The answer is related to   difference in temperature between original AQN with  $v_{\rm AQN}\sim 10^{-3} c$ and $AQN_s$ which could slow down to much lower velocities leading to much lower internal temperatures. The dramatic differences in temperatures between  parent AQN and daughter $AQN_s$ implies that the mean free paths
      for these cases are very different.  Furthermore, the AQN travels much longer distance in comparison with  $AQN_s$ during the same time interval. Therefore, the emissivity of the secondary photons  per unit volume for  these two cases will be  also very different. Therefore, all sky camera sensitivity might not be sufficient to record a weak  signal in visible bands from an  object moving with velocity $v_{\rm AQN}\sim 10^{-3} c$. However, IR emission recoding the heat should be strong in all cases. The observation of vapour trails in skyquakes events support this interpretation.}.

 A very mysterious  event  occurred on July 31-st 2008 was properly recorded by the dedicated Elginfield Infrasound Array (ELFO) near London, Ontario, Canada, where it was identified as the AQN-induced event, see   \cite{Budker:2020mqk}
for references and details.  The infrasound detection was accompanied by non-observation of any meteors by an all-sky camera network, ruling out a conventional meteor source. In addition, no any meteorites had been found in the active area, which is expected outcome for any AQN induced events, 
similar to ``Green fireballs" and  ``Elma" events mentioned above. 
%Anthropogenic sources such as operations at the nearby Bruce Nuclear Power Plant or the  Goderich salt mine were also eliminated; a  local airport radar reported no aircraft in the area at the time. 
%Furthermore, these events cannot be explained by military aircraft as  records of skyquakes appeared long before supersonic flights.  
In addition to infrasound, impulses were also observed seismically (few moments after the event) as ground-coupled acoustic waves around Southwestern Ontario and Northern Michigan. The seismic signal is also expected  as the AQN continues to propagate deep  underground as mentioned at the end of Sect.\ref{kinematics} in item 3. 
  This identification of ELFO event with AQN induced event is further supported 
  in \cite{Budker:2020mqk}
  where it was argued that    the energetics, infrasound-frequency properties, and other characteristics 
 of the event are consistent with observed ELFO event.

 \subsection{transmedia objects (water-air interface)}\label{sect:ANITA}
  
 It has been reported on number of  the  UAP events when object entering the water, or leaving the water and raising into  the air.   There are also  UAP events when the objects remain below the water surface and floating on the surface, see section 7 in   \cite{knuth2025newscienceunidentifiedaerospaceundersea} for references and details. Our comment here is that  in case when AQN hits the  water's surface it will continue to propagate with the same initial velocity  
  without producing  any splashes,  turbulence, or waves, as mentioned at the end of Sect.\ref{kinematics} in item 2.  The basic reason for such counter-intuitive behaviour
   is the same as before and  expressed by (\ref{image}). A naive  expectation of  splashes and waves 
    is related to a common misconception as summarized at the end of Sect. \ref{differences} and also  in footnote \ref{misconception}.

 It is quite  remarkable  that the events when UAP leaving the water and raising into  the air had been also observed. What  could it be in AQN framework? 
 In previous subsection we discussed ELFO event which was properly recorded by  dedicated  infrasound instrument. It was identified as AQN-induced  downward going event.  We also mentioned that with a typical for DM velocity $v_{\rm AQN}\sim 10^{-3} c$ the AQN propagates through atmosphere and Earth's core without loosing much of momentum and baryon charge for $B\gtrsim 10^{25}$.  It is tempting to assume that the   upward going UAP events from water, which apparently also had been observed,   see section 7 in   \cite{knuth2025newscienceunidentifiedaerospaceundersea}, could be related to the AQN-induced upward going events. 

In fact, we had previously suggested that similar  upward going AQN -induced events (ice -air rather than water-air) had been recorded in the radio frequency bands 
by  the Antarctic Impulse Transient Antenna (\textsc{ANITA})  collaboration.  To be more concrete, the \textsc{ANITA}  collaboration  \cite{Gorham:2016zah,Gorham:2018ydl,Gorham:2020zne} has reported two anomalous events with non-inverted polarity (upward going) events. 
These events correspond to very large inclination angle  when ``something" crossing the Earth before exiting from opposite side of Earth  at the moment of  recording by \textsc{ANITA}. Such events  are very hard to explain within conventional physics\footnote{Initially, people thought it could be $\nu_{\tau}$. It was ruled out later, see  \cite{Liang:2021rnv} for references and details about this and other proposed mechanisms to explain anomalous events recorded by \textsc{ANITA}.}, but could be  explained within AQN framework \cite{Liang:2021rnv}, including the energetic and spectral properties of the recorded radiation.  The estimation for the event rate for   anomalous events   with non-inverted polarity  reported by \textsc{ANITA} 
   is also consistent with   our formula for the flux   (\ref{Phi1}) within AQN framework.

\subsection{Other common UAP's features from AQN perspectives }\label{sect:others} 
There are many  observed features of the UAP events which could be related  to the AQN events as mentioned in previous subsections. In addition to those properties 
there are some  UAP reports with   the following puzzling features: 

1.   The objects may appear in group.  This is not an  unusual  element within  AQN proposal (\ref{identification}). Indeed, the multiple MT events grouped together    were observed  in \cite{Villarroel:2021ddh,Villarroel_2025,Villarroel:2025xym} and explained within AQN framework previously in \cite{Zhitnitsky:2026pai}. At the same time, the MT events and  the UAP events represent two different stages in evolution   of the same system. Therefore, the multiple UAP events could also occur. 

2. The UAP leaves a faint trail which has been also discussed previously in context of the ELFO events. The presence of vapour is expected and even predicted by the proposal (\ref{identification}),  see text  after eq. (\ref{lambda_AQN}).

3. The UAPs may change direction with enormous acceleration. The AQN$_s$ indeed may  change the direction of the motion if sufficiently strong atmospheric electric field is present in the area as discussed in Sect.\ref{E-field}. The time which is required to change the direction is measured in seconds, while typical distance for such maneuver  is   around 100 meters (could be much shorter  for lower velocity) according to (\ref{l}). The acceleration  AQN$_s$  experiences during such maneuver could be $10^2$g  according to estimate (\ref{a}), which is smaller than a typical deceleration $(10^2-10^4)$g experienced by AQN$_s$ due to the friction forces (\ref{eq:acceleration}) resulting from scattering and annihilation events. 

4. The fact that the UAPs moving with enormous velocity exceeding 5 times the speed of sound $c_s$ (or even more) is very generic feature of the AQN-induced events. For typical sizes $B\lesssim10^{25}$ these events do not produce any sound signals which can be heard by humans as discussed in Sect.\ref{acoustics}. Dedicated  infrasound instruments such as ELFO, see subsection \ref{sect:ELFO}  may detect infrasound signals for sufficiently large and very rare AQN events with $B\gtrsim 10^{27}$. 
  This  expectation to hear strong sound signal from fast moving body 
    is related to a common misconception as summarized at the end of Sect. \ref{differences} and    
    basically  expressed by  (\ref{image}) illustrating the  fundamental difference between the observed size of the objects and real physical size of the objects. 
    
    5. A typical AQN-induced event must have a form of a sphere (it could also assume a prolongated shape due to a high velocity) because the apparent size of the object is determined by the mean free path of X rays as explained  in Sect.\ref{visible-size}. The X ray emission from the nuclear core  normally proceeds in spherically symmetric way forming a plasma fireball.  The  universal formula (\ref{lambda1}) describes the  apparent  size of the object (in visible and IR), and it is  applicable to all kind of AQN-induced events, including the BL and MT events.  
    The real physical size of the core   is many orders of magnitude smaller as  it is measured in $\mu m$ rather in  meters, as we already mentioned. Precisely this  common    misconception    between apparent size of the object and real physical size of the same object leads to  
   many  puzzling features of observed phenomena, see last paragraph at the end of Sect.  \ref{differences} with  more comments.

We conclude this section with the following remark. We reviewed a small class of known UAP events with some specific features which match the expectations from AQN-induced events as described  in Sect. \ref{differences}. The required instruments which could  detect a signal from such kind of the UAP events will be discussed in next section. 
    There are many other UAP events (which do not belong to the considered  class) and which may have very different characteristics from those listed  above,  and which require  dramatically different type of instruments to study them, see e.g. the Galileo Project \cite{doi:10.1142/S2251171723400068} and UAPx project \cite{Szydagis:2024eea}.

\section {Concluding comments and Future Developments} \label{conclusion}

             The key element of the proposal (\ref{identification}) is   a  new  DM paradigm  when DM is represented by   AQN, is very strongly interacting object made of quarks, antiquarks  and gluons of SM particle  physics as reviewed in Sect. \ref{AQN}.                The AQN is a proper DM candidate in empty space. However it may produce a number of profoundly energetic events when AQN enters the Earth or other planets and stars. This is because the nuclear antimatter  material of the AQN's core  represents the fuel  which  powers  such  energetic events.

                   The UAP events are precisely  such kind of profound manifestations, among many, many others, including the  BL events, MT events, sky-quakes, pseudo-meteorites.  It is important to emphasize that all parameters which have been used in our estimates in key sections  \ref{differences} and   \ref{items} are not ad hoc parameters to fit the observations. Instead, the same parameters have been used previously to study  
                other puzzling cosmological observations at dramatically different scales (from early Universe, to galactic, to solar scale) and which could be naturally explained within   the same AQN framework, as briefly  reviewed in  
 \cite{Zhitnitsky:2021iwg}, discussed in lectures in \cite{VanWaerbeke:2026sxs} and explained in plain simplified language with illustrations in \cite{QCD-cosmology}. The basic consequence of this new DM paradigm in context of the UAP events is summarized by the last paragraph at the end of Sect. \ref{differences} (with a bullet sign), and we shall not repeat it here. Instead,  
in next few paragraphs we elaborate on  possible future search directions of the UAP events based on our understanding of a specific  class of the UAP events which are AQN-induced events as described in Sect. \ref{items}. 

 First of all, one can    install all sky cameras, similar to the ones used in  analysis of the BL spectrum \cite{BL-spectrum-1}, to monitor entire  sky.  
As we discussed above the BL events discussed in \cite{Zhitnitsky:2025bvy} within AQN framework  are   close cousins of the UAP events.   Therefore, we expect conventional  spectral  lines corresponding to the chemical elements present in atmosphere and dust from soil. 

  Another suggestion is to search UAPs in radio frequency bands. This suggestion is based on observation of two mysterious events recorded by  \textsc{ANITA}  as described in 
  ``trans-media objects" in Sect. \ref{sect:ANITA}. These puzzling events cannot be explained as a  conventional Cosmic Rays (CR) event, but could be explained within AQN framework  as argued in  \cite{Liang:2021rnv}. 
  
  Another puzzling observation we want to mention here    could be also related to the  AQN-induced event. Indeed,    the Telescope Array  (TA) collaboration  observed   the so-called  mysterious   bursts \cite{Abbasi:2017rvx,Okuda_2019}. The mystery here is that at least three air showers  were recorded within 1 ms (multiple simultaneous events representing the cluster) which cannot occur as a result of  conventional high energy CR event.   
                We suggested   in   \cite{Zhitnitsky:2020shd,Liang:2021wjx} that these puzzling  events  could be related to the AQN-induced events. What is important for the present work is that these events could be studied in radio frequency bands, and these events could be easily discriminated from the radio noise and spurious events as argued in  \cite{Liang:2021wjx}.
     The analysis of the time correlations between all sky camera observing  UAP  events and radio signals would unambiguously identify the source and nature of these events. 

One more suggestion on possible future UAP studies is based on observation that some BL events and pseudo-meteorites events had shown some signs of  hard radiation which always accompanies the events.  It may leave some evidences, see item 4 in Sect.\ref{kinematics} and item 6 in Sect.\ref{green_fireball}.    For example, one could measure the level of thermoluminescence of local minerals in comparison with the background level even many years  after the event, see \cite{Lazanu:2024hkk}. 
  Such studies are possible if the location of UAP hitting the surface is found, similar to ELMA event as described in Sect.\ref{sect:ELMA}.  Therefore, it would be amazing discovery  if some sort of signals of hard radiation could be  measured in future UAP studies, as it is very difficult (if even  possible)  to imagine any other feasible  sources for hard radiation for UAP.  
  %Such investigation    could be only done if affected area is found and localized. 
  Such an   analysis  can be  carried out in addition to studies   recently advocated   in 
  \cite{doi:10.1142/S2251171723400068} in context of the Galileo Project and in  \cite{Szydagis:2024eea} in context of UAPx project.

     We want to add few lines on estimation of the UAP event rate. The frequency of the   BL events has been estimated in          \cite{Zhitnitsky:2025bvy} within AQN perspective, and it has been shown to be consistent with observations. Important point here is that the estimations for the event rate were based on DM flux (\ref{Phi1}) and determined by the DM density.  
  Furthermore, the frequency of mysterious \textsc{ANITA}  events was also estimated from  the same formula (\ref{Phi1}), and it is consistent with observed anomalous events which we interpreted as AQN-induced events.  The recorded  frequency of appearance of the   TA mysterious   bursts (10 events recorded during 5 years)
   is also consistent with our estimation   based on  our normalization formula for the flux   (\ref{Phi1}).   We predict that the event rate for the UAP events must also be described by the same formula  for the flux (\ref{Phi1}) if our identification  (\ref{identification}) on the nature of some UAP events is adopted. This is because we consider the BL events,  \textsc{ANITA}  anomalous events,
   the TA mysterious   bursts are the AQN-induced events. As a result, the corresponding rate must be proportional to the same    flux (\ref{Phi1}) in terms of the DM parameters.

  The AQN dark matter model   is  consistent with all presently available cosmological, astrophysical, satellite and ground-based observations. The key player of this new DM paradigm is 
  the antimatter nuggets with typical characteristics (\ref{eq:values}). We emphasize that this model   was not introduced to explain the UAP events, which is the topic of the present work.  Rather, it was suggested long ago  \cite{Zhitnitsky:2002qa}  to resolve  two fundamental  cosmological puzzles: 1. similarity  between visible and DM components, $\Omega_{\rm DM}\sim \Omega_{\rm visible}$; 2. observed baryon asymmetry of our Universe. These puzzles are naturally  resolved within the  AQN framework without fitting any parameters. The application of this framework to UAP events as presented in Sect. \ref{differences} and Sect. \ref{items}  is a byproduct of the AQN construction. 
  
The  applications of  the same model to other systems 
      may even shed some light  on the  long standing puzzles and mysteries     in cosmos as reviewed in   \cite{Zhitnitsky:2021iwg,VanWaerbeke:2026sxs}.     If validated through the proposed tests and experiments on Earth as suggested above it would unambiguously identify the nature of the DM and its role in cosmos and Earth, and relate it to some  UAP events observed for decades without clear physical explanation of the events, similar to BL and skyquakes (being UAP cousins)  recorded for centuries. One can say that the future UAP studies   could play a key role in the {\it direct } studies  of the DM physics, in contrast with
{\it indirect}  manifestations of the AQN-induced phenomena at different cosmological scales, see brief   review    
 \cite{Zhitnitsky:2021iwg},  the  lectures  \cite{VanWaerbeke:2026sxs} delivered at the Les Houches Summer School  for MSc and PhD students, and the video (with illustrations) \cite{QCD-cosmology} for BSc students.

     \section*{Acknowledgements}
      This research was supported in part by the Natural Sciences and Engineering
Research Council of Canada.

  \bibliography{UAP.bib}

\end{document}